\documentclass[12pt]{article}
\usepackage[applemac]{inputenc}\usepackage[english]{layout}
\usepackage[english]{babel}
\usepackage{multicol}
\usepackage{ae,aecompl}
\usepackage{fancyhdr}
\usepackage{amsmath,amssymb,latexsym, cancel, feyn}
\usepackage{amssymb,latexsym}
\usepackage{graphicx}
\usepackage[pdftex]{color}
\usepackage{xcolor}
\usepackage{soul}
\usepackage{bbold}
\usepackage[small,sf]{caption}
\usepackage[headheight=10pt,headsep=15pt,footskip=1.5cm,hmargin=2.0cm,includeheadfoot,top=2cm,bottom=1.5cm]{geometry}
\usepackage{hyperref}

\title{Photon emission in QGP using AdS/QCD at finite chemical potential}
\author{M. A. Martin Contreras\thanks{ma.martin41@uniandes.edu.co}, J. M. R. Roldan Giraldo\thanks{jroldan@uniandes.edu.co}\\ \small High Energy Group, Department of Physics, Universidad de los Andes}
\date{}
\begin{document}
\maketitle
\begin{abstract}
We calculate the photon emission rate and the electrical conductivity of the QGP at finite temperature and finite chemical potential using AdS/QCD approximations in an AdS Reissner Nordstrom background. To do so, we supposed the medium properties to be encoded in a geometric background. The results obtained in the hard wall and soft wall model are consistent with the observed phenomenology and they also in agree with other holographic results, as the D3/D7 or the Sakai Sugimoto models, suggesting the universality of AdS/CFT conjecture as tool to explore QCD. 
\end{abstract}

\section{Introduction} 
In its beginnings, string theory was expected to be a possible model for strong interactions. But it was eclipsed by the greatest QCD outcomes, making it to be forgotten as fundamental theory for such interactions. Nowadays, with the development of the so-called gravity/gauge duality, string theory recovers its original motivation becoming a tool to explore the strong interactions at their non-perturbative region even at extreme scenarios with finite temperature and chemical potential. These scenarios are observed at nuclei accelerators and stars center. Duality itself suggests that gravitational dual of QCD should exist. The real problem is how to find it.\par  

In this direction, the most successful approximation to gravity/gauge duality is the AdS/CFT Correspondence by J. Maldacena \cite{Maldacena-1}, which has been used as a tool to describe non-perturbative hadronic phenomena, specially the quark gluon plasma observed in heavy ion collisions. Properties such as the entropy density, bulk viscosity, jet quenching, photon/dilepton emission rate were well described holographically  in the zero chemical limit by top/down approaches as the Dp/Dq system \cite{Mateos-2} or the Sakai Sugimoto model \cite{Bu-2}. Also the bottom/up realizations as the improved holographic QCD model \cite{Kiritsis-1} have described the QGP system.  Recently, \cite{Yang-1} have discused the shear viscosity and photon emission in the QGP in the latter model. In the AdS/QCD realm, some approximations have been made in the references \cite{Nata-1} where they calculated the spectral density for photons in a specific gravitational setup using the soft wall model. Also \cite{Bu-1} has calculations for the hard wall and the soft wall model for finite chemical potential modeled by the probe D-brane method, in which the DBI action is minimized in order to obtain the embedding profile generated by a U$\left(1\right)$ static field.  A good review of the holographic approximation to the QGP phenomenology is \cite{Janik2011}.\par

This paper is organized as follows: in section 1 we discuss the photon emission rate and the conductivity from the quantum field theory side. In section 2, we set the general background for both hard wall and soft wall models. In section 4 it is discussed how we calculate the trace of the spectral function analytically for each AdS/QCD model. In section 5, we consider the numerical results obtained for the photon emission rate and the conductivity. Finally in section 6 we enunciate the conclusions for our work. \par
\section{Photon Emission Rate} 
It is well known that any thermal medium composed of charged particles will radiate photons. This photon spectrum can be used as a probe to study the characteristics of such thermal medium, since it depends on its specific properties.\par

In the case of the QGP produced in heavy ion collisions, the thermal medium is optically transparent due to the small spatial extension\footnote{At this point, the assumption that the photon mean free path is bigger than the plasma ball size was taken.} and the smallness of electromagnetic coupling. Therefore, any emitted photon escapes from the plasma ball without any further interaction with the medium. In such a situation the spectrum can be considered as \emph{almost planckian}, but it is still useful to obtain information about the plasma in the emission point.\par

At this point a remark must be done: depending on the momentum $k$ of the emitted photons the dynamics can be split in specific regimes defined by three characteristic length scales, the hard, soft and ultra-soft scales \cite{Shuryak}.\par

The hard scale is defined when the limit $k\approx T$ is reached. Contributions of excitations in this scale dominate the bulk thermodynamic properties such as total momentum density and total energy of the plasma, likewise conserved charge susceptibilities. These excitations also control the transport properties in the colored medium.\par

The second scale, soft scale, is reached when $k\approx g T$, where $g$ is the gauge theory coupling. In this scenario the Landau damping and also the Debye screening, due principally to the coherent interaction of the particles in the thermal colored medium, become important. These effects are treated with the hard-thermal-loop perturbation theory \cite{Singh-1,Gervais-1,Arnold-1,Arnold-2}. \par

The third scale is the ultrasoft or non-perturbative: $k\approx g^2T$. In this scale, low frequency and long wave-length dynamics gives the hydrodynamical regime. Photons produced in this regime carry information about properties such as the electrical conductivity. Soft and hard photons produced in more energetic scenarios tends to be suppressed due to the Landau Pomeranchuk Midgal effect. Photon production in a colored medium was discussed in detail both from strong and weak coupling point of view in reference \cite{Huot-1}.\par

Let us consider a QGP defined by a colored  quantum system in equilibrium that can be described by a thermal field theory in the strongly coupling limit.  It will be assumed that the electromagnetic interaction between matter and photons is given by the minimal coupling of a photon field with the electromagnetic current, $e\, A^{\mu}J_{\mu}^{em}$, with $e$ the electromagnetic coupling. \par

At leading order, the photo-production is given by the expression \cite{Arnold-1}
\begin{equation}\label{photo_rate}
d\Gamma_{\gamma} = \left.\frac{d^3q}{\left(2\pi\right)^3}\,\frac{e^2}{2 \left(k^0\right)^2}\,\eta^{\mu\nu}\,C_{\mu\nu}^R\left(k\right)\right|_{k^0=|\vec{q}|},
\end{equation}
where we have defined the photon 4-momentum as $k^{\mu}=\left(k^0,\vec{q}\right)$, $\eta_{\mu \nu}$ denotes the Minkowski metric and $C_{\mu \nu}^R\left(k\right)$ is the Wightman function defined for electromagnetic currents as,

\begin{equation}\label{wightman}
C^R_{\mu\nu}\left(k\right)=\int{d^4x\,e^{-ik x}\,\langle J^{EM}_{\mu}\left(0\right)\,J^{EM}_{\nu}\left(x\right)\rangle}.
\end{equation}

Notice that the expectation value in (\ref{wightman}) is taken in the thermal equilibrium state and $x^{\mu}=\left(x^0,\vec{x}\right)$. In the thermal equilibrium limit the Wightman correlator can be written in terms of the spectral density, 

\begin{eqnarray}\label{spectral1}
C^R_{\mu \nu}&=&n_B\left(k^0\right)\,\chi_{\mu\nu}\left(k\right),\\ \label{spectral2}
\chi_{\mu\nu}&=&-2\,\mathbb{Im}\,G^R_{\mu\nu}\left(k\right),
\end{eqnarray}

with $n_B$ the Bose-Einstein distribution and $G^R$ the thermal retarded Green function. At non zero temperature, Lorentz invariance is explicitly broken leaving us a residual rotational symmetry that allows, together with the gauge invariance, to split the thermal Green function into longitudinal and transversal parts.\par

Since the idea is to model real photon production (implying $k^0=|\vec{q}|$), it is possible to ignore the longitudinal part and focus on the transversal one only. Then, the trace of the spectral function (\ref{spectral2}) is written as

\begin{equation}\label{spectral3}
\chi^{\mu}_{\mu}\left(k\right)=-4\,\mathbb{Im}\,\Pi^T\left(k\right), 
\end{equation}

where $\Pi^T$ is the transversal part of the spectral density.  Keeping in mind all these definitions and results, the photon emission rate is given by 

\begin{equation}\label{photonrate}
\frac{d\Gamma_{\gamma}}{d^3|\vec{q}|}=\left.\frac{e^2}{\left(2\pi\right)^3\,2|\vec{q}|}\,n_B\left(k^0\right)\,\chi^{\mu}_{\mu}\left(k\right)\right|_{k^0=|\vec{q}|}.
\end{equation}

The electrical AC and DC conductivities of the medium are also determined by the trace of the spectral function. From the Kubo formula the AC $\sigma\left(k^0\right)$ conductivity can be read from the spatial components of $\chi_{\mu \nu}$ as 

\begin{equation}	\label{ACconduc}
\sigma_{\text{AC}}\left(k^0\right)=-\frac{\chi_{ii}\left(k^0,\vec{k}=0\right)}{2ik^0}.
\end{equation}

From the spectral density in the limit of $k_0 \rightarrow 0$ we can read the DC conductivity as

\begin{equation}\label{DCconduc}
\sigma_{\text{DC}}=\left.\frac{\alpha_{\text{em}}\pi}{2 T}\frac{d\chi^{\mu}_{\mu}\left(k^0\right)}{d k^0}\right|_{k^0=0}.
\end{equation}

Then the problem of calculating the photon emission rate and the conductivity translates into the calculation of the trace of the spectral function holographically. This procedure will be realized in the next sections.\par

\section{Holographic Setup}
\begin{subsection}{Geometric Background}
We will focus on the calculation of the 2-point function using two of the AdS/QCD models: the hard wall \cite{Polchinski-1,Braga-3} and the soft wall \cite{Son-2} models. These two approximations can be summarized into a single expression just by considering the behavior of a statical dilaton $\Phi$ field.\par

In both AdS/QCD models the background is considered dual to the colored medium properties, composed by massless quarks. Since we are considering the finite chemical potential case, it is possible to introduce a charge in the AdS background through an Einstein--Hilbert--Maxwell action. Holographically, we will say that this charge is dual to the generator of the  $U\left(1\right)$ baryonic symmetry.  Following these ideas the background action can be written as

\begin{equation}\label{bg_action}
I_{\text{Background}}=\frac{1}{16\,\pi G_5}\int{d^5x\,\sqrt{-g}\,e^{-\Phi}\left(R-2\Lambda-\frac{1}{4 g_5^2}G_{mn}G^{mn} \right)},
\end{equation}

where $G_{mn}:=\partial_m\,V_n\left(z,x^{\mu}\right)-\partial_n\,V_m\left(z,x^{\mu}\right)$, the cosmological constant is $\Lambda=-\frac{6}{L^2}$, with $L$ the AdS radius, $G_5$ is the Newton constant in five dimensions,and the five dimensional gauge coupling is defined as $\frac{1}{g_5^2}=\frac{N_c\,N_f}{4\pi}$, with $N_c$ and $N_f$ the number of colors and flavors respectively. Adding temperature to a given system holographically means the presence of a black hole, so it is possible to conclude that the minimal coupled Hilbert--Einstein--Maxwell  action (\ref{bg_action}) leads to a the Reissner--Nordstrom AdS Black Hole solution. In other words, this RN AdS BH solution can be considered as the gravitational dual for a strongly coupled plasma under finite chemical potential and finite temperature conditions \cite{Jo-1}. Following these ideas, the background is written as:
 
\begin{equation}\label{metric_RN_AdS}
dS^2=\frac{L^2}{R^2}\left[-h\left(z\right)dt^2+d\vec{x}^2+\frac{dz^2}{h\left(z\right)}\right],
\end{equation}
   
where the blackening factor $h\left(z\right)$ is fixed by the 1-form potential $V_m$. If a static solution is chosen, $V_m=-V\left(z\right)\,dt$,  the $h$ factor can be written as 

\begin{equation}\label{blackening}
h\left(z\right)=1-(1+q^2z_h^6)\left(\frac{z}{z_h}\right)^4+q^2z_h^6\left(\frac{z}{z_h}\right)^6,
\end{equation}

where $q$ is the charge of the black hole and $z_h$ is the event horizon. The temperature is defined using the Hawking--Page relation \cite{Hawking-1,Cai-1}

\begin{equation}\label{temp}
T=\frac{\left|h'\left(z_h\right)\right|}{4\pi}=\frac{1}{\pi\,z_h}\left(1-\frac{q^2\,z_h^6}{2}\right).
\end{equation}

In pure AdS, i.e. without dilaton, the 1-fom potential is given by $V\left(z\right)=\mu-q^3z_h^3 z^2$. In AdS/QCD models the same form cannot be granted due to the presence of the cutoff\footnote{The case of the Hard Wall model is similar to pure AdS since the dilaton is fixed to be zero, but the AdS space is sliced. If it is possible to ensure that the sliced part of the geometry lies outside of the RN AdS BH then the 1-form potential is the same as in pure AdS.}.  In that case it is possible to approximate the AdS/QCD solution to the pure AdS one in the limit $z\rightarrow 0$ \cite{Colangelo-2}. Holographically, the chemical potential $\mu$ related to the gauge theory in the boundary is defined in the limit $z\rightarrow 0$ where the 1-form static potential is $V(0)=\mu$.
\end{subsection}

\begin{subsection}{Holographic Photons}
The QGP will be modeled using a 5-dimensional $U\left(1\right)$ field $A_m$ coupled to the static dilaton in the AdS/QCD standard form
\begin{equation}\label{photons}
I_{\gamma}=-\frac{1}{4g_{\gamma}^2}\int{d^4x\,\sqrt{-g}\,e^{-\Phi\left(z\right)}F_{mn}F^{mn}},
\end{equation}

where $g_{\gamma}^2$ is the coupling constant for photons in the QGP, that in general depends on temperature \cite{Patino-1,Nata-1} and $F_{mn}=\partial_m \,A_n-\partial_n\,A_m$.  From the photon action it is possible to obtain the equations of motion for the vector field $A_m$. To do this, the photons will be supposed to be moving in the $x_3$ Euclidean direction, then the photon 4--momentum can be fixed to be $k_{\mu}=\left(k_0,0,0,k_0\right)$. Therefore, using the background (\ref{metric_RN_AdS}) the e.o.m. in Fourier space are written as

\begin{equation}\label{eqn_mov}
\partial_z \left[\frac{e^{-\Phi\left(z\right)}}{z}h\left(z\right)\partial_z A\left(k,z\right)\right]+\frac{e^{-\Phi\left(z\right)}}{z}k_0^2\left[1-h\left(z\right)\right]A\left(k,z\right)=0,
\end{equation}

where the gauge fixing $A_z=0$ was implicitly used and $A\left(q,z\right)$ is  the transversal field component. The longitudinal components, related to the dilepton production are fixed to be zero. 
\end{subsection} 

\begin{subsection}{Thermal Green Function}

The retarded Green function $G^R$ can be constructed with the solutions of eqn. (\ref{eqn_mov}).  In order to do that, the first step is to evaluate the on-shell boundary photon action obtained from (\ref{photons}):

\begin{equation}
I_{\text{On--Shell}}^{\text{Bndry}}=-\frac{L}{2g_{\gamma}^2}\int{d^4k\,\left[\frac{e^{-\Phi\left(z\right)}}{z}h\left(z\right)A\left(z,k\right)\partial_z A\left(z,-k\right)\right]_{z\rightarrow 0}^{z=z_h}}.
\end{equation}

Following the Minkowskian prescription given in \cite{Starinets-1}, the thermal Green function can be obtained from the integrand in the boundary action, evaluated in the event horizon $z_h$:
\begin{equation}
G^R\left(k\right)=\frac{1}{g_{\gamma}^2}\left.\frac{e^{-\Phi\left(z\right)}}{z}h\left(z\right)A\left(z,k\right)\partial_z A\left(z,-k\right)\right|_{z=z_h}.
\end{equation}

With this expression, the trace of the spectral function (\ref{spectral2}) is given by  

\begin{equation}
\chi_{\mu}^{\mu}\left(k\right)=-\left.4\,\frac{e^{-\Phi\left(z\right)}}{z}h\left(z\right)\,\mathbb{Im}\,A\left(z,k\right)\partial_z A^*\left(z,k\right)\right|_{z=z_h}
\end{equation}
\end{subsection}

The trace of the spectral function is calculated from the action (\ref{photons}) using the static background (\ref{metric_RN_AdS}). In the next sections the discussion will be focused in each AdS/QCD approximation, namely, hard wall and soft wall models. \par

\section{AdS/QCD models}
\begin{subsection}{Hard wall model}
Hard wall is defined by fixing the dilaton field to be zero, $\Phi\left(z\right)=0$, together with  the Dirichlet condition on the transverse fields at some holographic  coordinate $z_{\Lambda}=1/\Lambda_{QCD}$ taken as a hard cutoff, where $\Lambda_{QCD}$ is the QCD energy scale.  The physical behavior of the hard wall system is equivalent as the square well in quantum mechanics: the presence of the cutoff induces a discrete set of solutions that defines the eigenstates of the system.\par 

The  background action in the hard model is then given by
\begin{equation}\label{HWMBG}
dS^2=\frac{L^2}{R^2}\,\Theta\left(z_{\Lambda}-z\right)\left[-h\left(z\right)dt^2+d\vec{x}^2+\frac{dz^2}{h\left(z\right)}\right],
\end{equation}
where $\Theta\left(z\right)$ is the usual step function that defines the AdS slice and $h$ is the warp factor defined in the expression (\ref{blackening}). The temperature in this case follows from (\ref{temp}).\par

To obtain the charge in terms of the chemical potential $\mu$ it is necessary to solve the equation of motion for $V\left(z\right)$ in the AdS slice and then evaluate the Hilbert--Einstein--Maxwell action to construct the gran canonical potential $\Omega\left(\mu, T\right)$.\par

In order to have a regular solution we will impose that $z_{\Lambda}=z_h$. From the Hilbert--Einstein--Maxwell action evaluated in the hard wall background (\ref{HWMBG}) it is possible write the equation of motion for the electrostatic potential $V$ as:

\begin{equation}
\partial_{z}\left[\frac{1}{z}\partial_z V\left(z\right)\right]=0,
\end{equation}

Subjected to the boundary conditions $V\left(z_{\Lambda}\right)=0$ and $V\left(0\right)=\mu$. The solution  under these conditions is $V\left(z\right)=\mu\left(1-\frac{z^2}{z_{\Lambda}^2}\right)$, and after evaluating the on--shell Hilbert--Einstein--Maxwell action, the grand canonical potential is written as \cite{Nakamura-1,Seo-1,Nakamura-2}: 
 
\begin{equation}
\Omega\left(\mu, T\right)=-\frac{\mu^2 \tau}{g^2_5\, z^2_{\Lambda}\left(T\right)},
\end{equation}

with $\tau$ a fixed volume in the usual Euclidian space than can be fixed to be one and $z_{\Lambda}$ is a function of $T$ . \par

From the classical thermodynamic relation

\begin{equation}\label{thermo}
q=-\left.\frac{\partial\,\Omega}{\partial\,\mu}\right|_{T},
\end{equation}

we find the relation between the baryonic charge and the chemical potential. Then, we obtain:

\begin{equation}
q=\frac{2 \mu \tau}{g^2_5\,z_{\Lambda}^2}.
\end{equation}

Now that the background is fully characterized we can fix our attention on the photon action in the AdS slice in order to obtain the trace of the spectral density $\chi^{\mu}_{\mu}$. The e.o.m. for the photon field in this hard wall scenario is

\begin{equation}\label{EoMHW}
\partial_z \left[\frac{1}{z}h\left(z\right)\partial_z A\left(z\right)\right]+\frac{k_0^2}{z\,h\left(z\right)}\left[1-h\left(z\right)\right]\,A\left(z\right)=0.
\end{equation} 

To solve this equation, it is necessary to set the incoming boundary condition as the field $A$ approaches to the horizon. To do this, the field $A$ will be written as  $A\left(z,k\right)=\left(1-\frac{z}{z_{\Lambda}}\right)^{-i\omega z_{\Lambda}/(2-Q^2)}F\left(k,z\right)$, where the $F$ function is fixed to be 1 in $z_{\Lambda}$. Following these ideas we are able to write the trace of the spectral density:

\begin{equation}
\chi_{\mu}^{\mu}\left(k^0\right)=-\frac{4\,\mathcal{N}\,k^0}{z_{\Lambda}}\frac{\left(1-q^2z_{\Lambda}^6/2\right)}{\left|A\left(0,k_0\right)\right|^2}
\end{equation}

where the coupling constant is defined as $1/g^2_{\gamma}=N_c N_f T^2 /8\pi^2\equiv \mathcal{N}$ \cite{Arnold-2}.  This coupling measures the electric degrees of freedom in the plasma and scales with the temperature, since it enhances the photon production as a consequence of growing the plasma ball.\par

In general, equations like (\ref{EoMHW}) have no analytical solutions  since their highly non-linear behavior, so we need to make use of numerical methods in build up the trace of the spectral densit.  In order to do this analysis we will define the adimensional frequency $\omega=k^0/2\pi T$ and baryonic density charge $Q=qz_{\Lambda}^3$. The results for the trace of the spectral function and the photon production are showed in the figure \ref{fig:figone}.\par

\begin{subsection}{Soft wall model}
The static background in the soft wall is defined by the dilaton profile $\Phi\left(z\right)= \kappa^2 z^2$, where $\kappa$ is an energy scale of the hadron spectra  in a given Regge trajectory.  In this case, it is not univocally defined in terms of the temperature and baryonic density. Thus, as a first approximation, the values of $Q=qz_h^3$ will be taken small in order to use the relation $T=0.4917\kappa^2$ given by \cite{Herzog-1}.\par
  
The chemical potential can be obtained from the Hilbert--Einstein--Maxwell action (\ref{bg_action}) in this background by virtue of the e.o.m. 
\begin{equation}
\partial_{z}\left(\frac{e^{-\kappa^2 z^2}}{z}\partial_z V\left(z\right)\right)=0. 
\end{equation}

With the boundary conditions $V\left(z\right)=\mu$ and  $V\left(z_h\right)=0$, the solution for the equation above is in the form 

\begin{equation}
V\left(z\right)=\frac{\mu}{1-e^{\kappa^2z_h^2}}\left(e^{-\kappa^2z^2}-e^{-\kappa^2z_h^2}\right),
\end{equation}

In order to calculate the baryonic charge we will impose that the 1-form static potential in the limit $z\rightarrow 0$ should match the RN AdS version of it \cite{Colangelo-2}. Evaluating the on-shell Hilbert--Einstein--Maxwell action and using the thermodynamic relation (\ref{thermo}) we obtain the baryonic charge as a function of the chemical potential:

\begin{equation}
q=\frac{2L\tau}{g_5^2}\frac{\mu \kappa^2 e^{\kappa^2z_h^2}}{e^{\kappa^2 z_h^2}-1}.
\end{equation}

\begin{figure}[t]
	\centering
		\includegraphics[width=1 \textwidth]{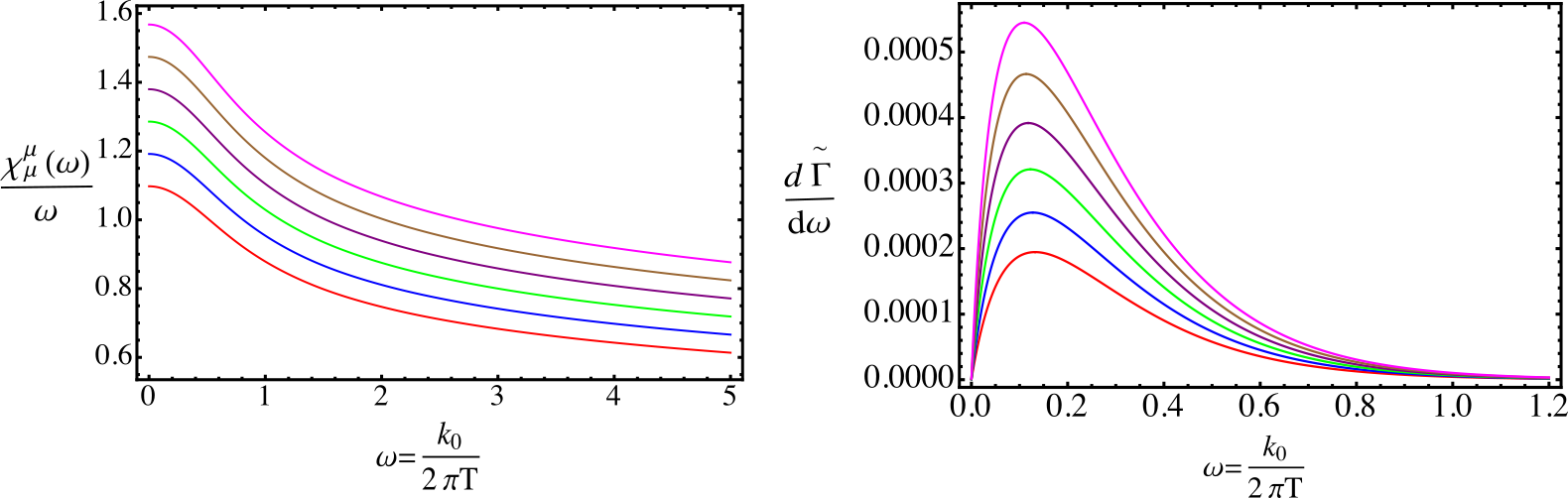}
	\caption{Left panel shows the trace of the normalized spectral function different values of $Q=qz_{\Lambda}^3$: $Q=0$ (red), $Q=0.1$ (blue), $Q=0.25$ (green), $Q=0.5$ (purple), $Q=0.75$ (brown), $Q=1.0$ (magenta); at the deconfinement temperature for this model given by $T_c=\frac{2^{1/4}}{\pi z_{\Lambda}}\left(1-\frac{Q^2}{2}\right)$  in the hard wall model \cite{Herzog-1}.  Right panel shows the normalized (divided by $\mathcal{N}$) photon emission rate at $T_c$ with the same values of $Q$.}
	\label{fig:figone}
\end{figure}

Once you have fully characterized the background, the next step is the construction of the spectral function  starting from the e.o.m in this case

\begin{equation}
\partial_z \left[\frac{e^{-\kappa^2z^2}}{z}h\left(z\right)\partial_z A\left(z\right)\right]+\frac{k_0^2\,e^{-\kappa^2z^2}}{z\,h\left(z\right)}\left[1-h\left(z\right)\right]\,A\left(z\right)=0.
\end{equation}

Imposing the ingoing boundary conditions as in the hard wall case we arrive to the trace of the spectral function given by the expression

\begin{equation}
\chi_{\mu}^{\mu}\left(k^0\right)=-\frac{4\,\mathcal{N}\,k_0\,e^{-\kappa^2z_h^2}}{z_{\Lambda}}\frac{\left(1-q^2z_{\Lambda}^6/2\right)}{\left|A\left(0,k_0\right)\right|^2}
\end{equation}

with the same $\mathcal{N}$ coupling as in the hard wall is used here.  Using the trace of the spectral function it is possible to construct numerically the photon emission rate in terms of $Q$ and $\omega$. The results in this approximation are plotted in the figure \ref{fig:figtwo}.
\end{subsection}

\section{Results}
\begin{figure}[h]
	\centering
	\includegraphics[width=1 \textwidth]{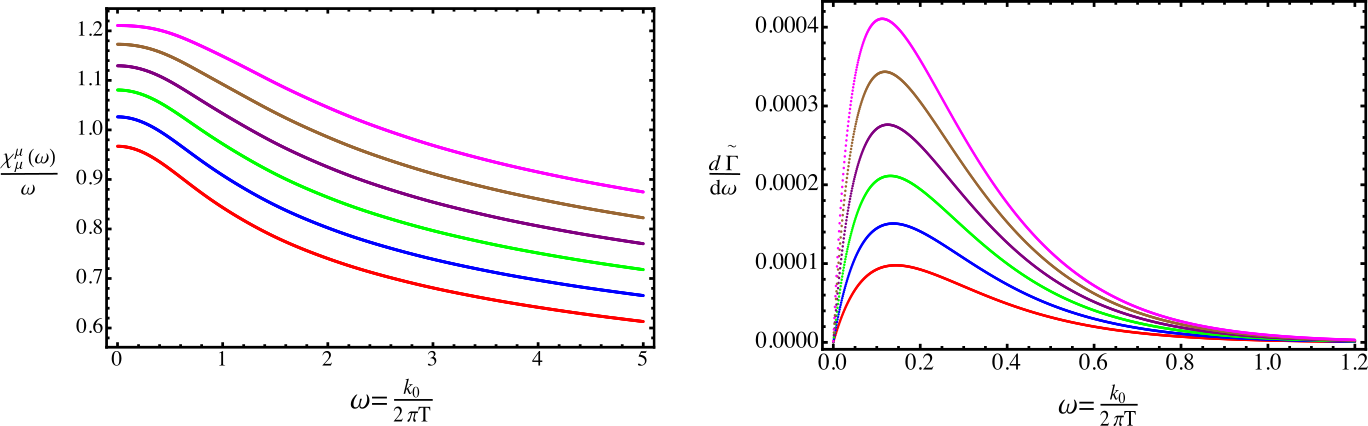}
	\caption{Left panel shows the trace of the normalized spectral function for different values of $Q=qz_{\Lambda}^3$: $Q=0$ (red), $Q=0.001$ (blue), $Q=0.0025$ (green), $Q=0.0050$ (purple), $Q=0.0075$ (brown), $Q=0.0100$ (magenta); at the deconfinement temperature for this model given by $T_c=0.4917\kappa^2$   in the soft wall model\cite{Herzog-1} for small values of Q.  Right panel shows the normalized photon emission rate at $T_c$ with the same values of $Q$.}
	\label{fig:figtwo}
\end{figure}

In order to do the numerical analysis in frequency, the temperature is fixed while the baryonic charge is varying. The results for hard wall and soft wall models are plotted in the figures \ref{fig:figone} and \ref{fig:figtwo}. 

The hard wall model is a holographic bag in the sense that confinement is placed by a hard cutoff fixed by the temperature of the colored medium. Any other extra properties of the medium are hidden in the background. For example, in Dp/Dq systems the quark mass is given by the embedding of the Dq--branes into the Dp--black branes in the decoupling limit \cite{Mateos-2}. In the case of AdS/QCD models, this specific information can be added with extra fields in the Hilbert--Einstein--Maxwell action. One application of this idea can be seen in \cite{Martin-1} where the quark mass is introduced as a scale defined by a bulk mass in a static modified hard wall. \par

In the case of the soft wall model, the cutoff is a decreasing function of the coordinate $z$ defined by a static dilaton. The energy scale $\kappa$ is fixed by the temperature of the medium. In other applications of the soft wall model to hadronic spectra the energy scale is fixed by the lightest particle in the spectrum \cite{Son-2}.  It is also possible to introduce massive quarks by adding extra fields into the background action. This will be done in future works. \par

The trace of spectral density on both models behaves monotonically with $\omega$ for the explored values of $Q$. The soft wall model is more sensitive to increasing of $Q$ than the hard wall. This is due to the presence of the dilaton field in the first one. The scale $\kappa$ is a function of $T$ and $Q$. For values of $Q$ near to the upper limit, $\sqrt{2}$, the trace of the spectral function is expected to begin to oscillate because the background is unstable for those values of $Q$ \cite{Cai-1}. The results for $Q=0$ are also in agreement with those obtained in \cite{Nata-1}, where the authors construct a different background in order to simulate the results for the D3/D7 system in the massless quark region and try to study how the trace of the spectral function behaves as a function of $\kappa$. Our background and the one showed in that reference differs by the transformation $u=\frac{z^2}{z_h^2}$. It is no surprising that these models have similar behaviors. In fact, Dp/Dq systems and AdS/QCD models are AdS-like (or conformal AdS) in the sense of \cite{Skenderis-4}.\par

\begin{figure}[t]
	\centering
		\includegraphics[width=1 \textwidth]{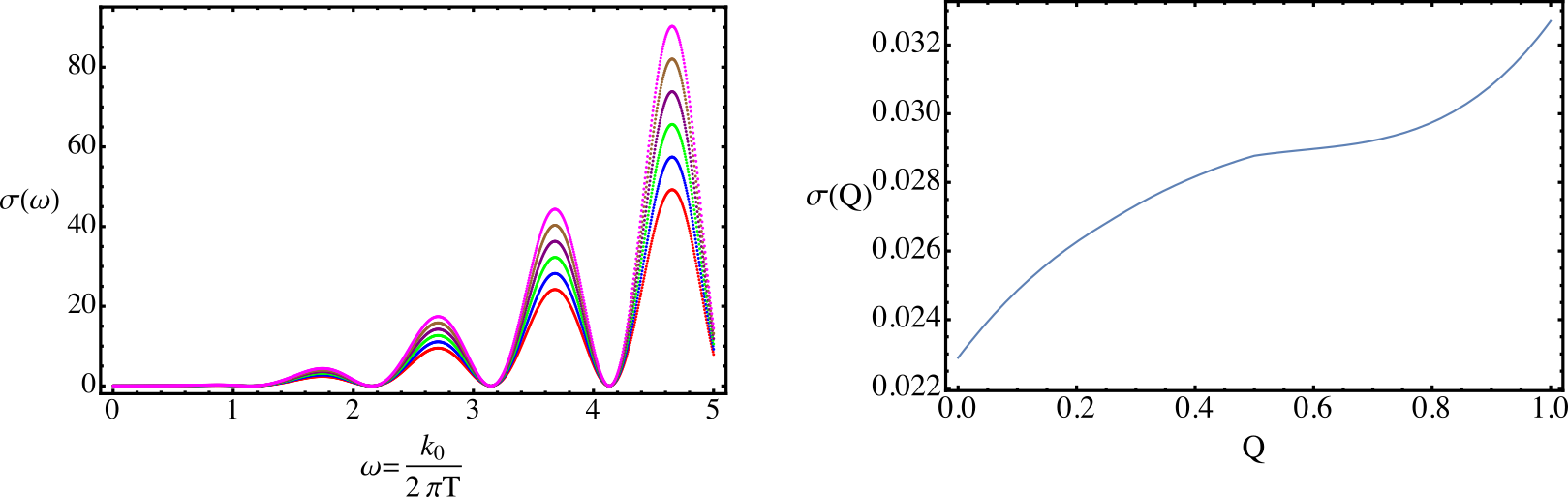}
	\caption{Left panel shows the real part of the AC conductivity calculated with the expression (\ref{ACconduc}) for the hard wall model as a function of the normalized frequency $\omega$ for the same values of $Q$ used in figure \ref{fig:figone}. Right panel shows the DC conductivity calculated from the equation (\ref{DCconduc}) in the hard wall model as a function of the baryonic charge $Q$.}
	\label{fig:figthree}
\end{figure}

The photon emission rate obtained for both models (right panels in figures \ref{fig:figone}
And \ref{fig:figtwo}) are also consistent with the QGP phenomenology. The emission occurs for small values of $\omega$, where the ultrasoft behavior is expected \cite{Endres-1}. When $\omega$ is increased the photon emission is damped, according to the Landau damping effect and Debye screening: photons are reabsorbed by the medium and are suppressed.  For high values of $\omega$, the emission goes to zero due to the reduction of bremsstrahlung and pair production, the so called Landau--Pomeranchuk--Migdal effect \cite{Dutta-1,Aurenche-1,Aurenche-2}.\par

When the chemical potential is turned on, the chemical equilibrium is broken, implying that there is an excess in the quark (or anti-quark) population that makes the enhancement of the quark (anti quark) bremsstrahlung over their annihilation radiation, causing that the photon emission rate increases with the chemical potential for low energies \cite{Singh-1}. For all the $Q$ values tested, the photon emission rates have common asymptotic behavior with large $\omega$, demonstrating the expected universality in 4+1 strongly coupled plasmas. These results are in agreement with reference \cite{Bu-1}, where the photon emission rate was calculated starting from the DBI action instead of considering an AdS RN black hole. Holographically, turning on the chemical potential means the phase transition from the AdS Schwarzschild to the AdS RN background.\par   

\begin{figure}[t]
	\centering
		\includegraphics[width=1 \textwidth]{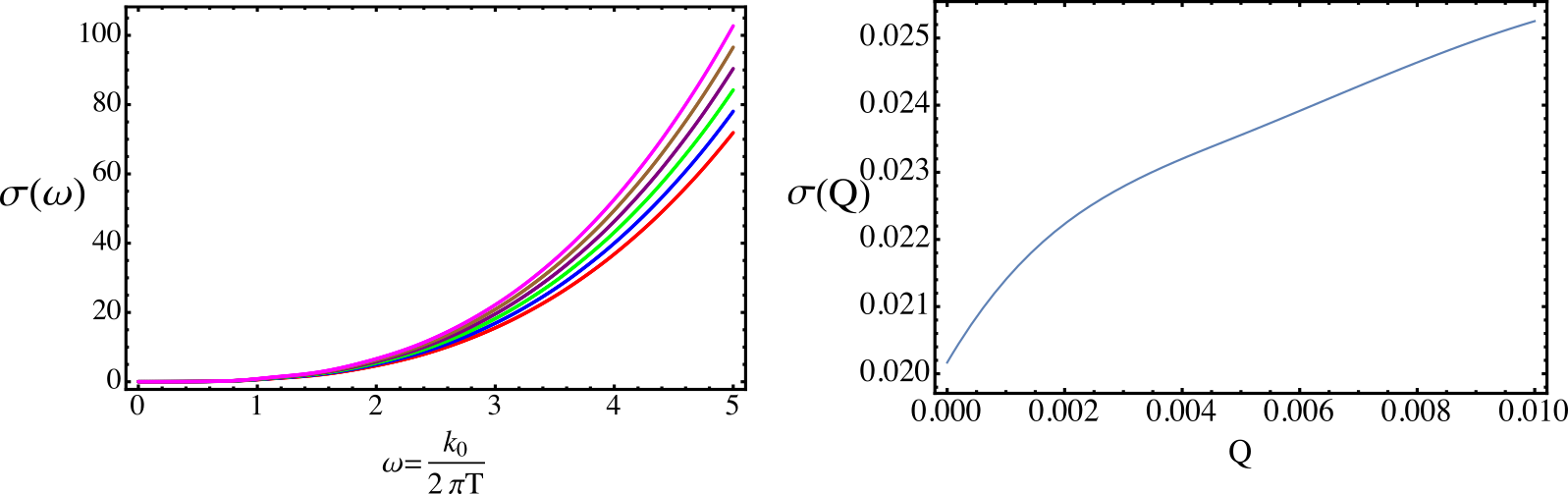}
	\caption{Left panel shows the real part of the AC conductivity calculated with the expression (\ref{ACconduc}) for the soft wall model as a function of the normalized frequency $\omega$for the same values of $Q$ used in figure \ref{fig:figtwo}.  Right panel shows the DC conductivity calculated from the equation (\ref{DCconduc}) in the soft wall model as a function of the baryonic charge $Q$.}
	\label{fig:figfour}
\end{figure}

In the figures \ref{fig:figthree} and \ref{fig:figfour} the AC and DC conductivities are plotted. In the hard wall case, the real part of the AC conductivity (left panel in both figures) grows monotonically with $\omega$ up to a maximum, then it starts to oscillate. In the case of the soft wall model, the dilaton prevents that the real part of the AC conductivity begins to oscillate, damping the profile.  In the case of the DC conductivity (right panels), for both models it is increasing with the baryonic charge for a fixed temperature. This is expected since in the limit $\omega \rightarrow 0$  we are in the ultrasoft limit, where the chemical potential enhances the  photon production by the reasons exposed above. 
\end{subsection}

\section{Conclusions}
In this work we have discussed the photon emission rate and the conductivity for hard wall and soft wall model under finite chemical potential regime.  We have constructed holographically the chemical potential by means of an AdS Reissner--Nordstrom black hole with a static electrical charge.  In the perspective of QFT, the holographic procedure showed here is equivalent to consider all the Feynman diagrams allowed by the Landau--Pomeranchuk--Midgal effect that have photons as external legs. \par

It is also interesting to comment that the results showed here are in agreement with other top/down models used to study photon emission \cite{Bu-1,Patino-1} despite the fact that they are constructed in different backgrounds, as in the case of the Sakai Sugimoto discussed in \cite{Bu-2}. The evidence that different holographic approaches can model the same phenomenology encodes an underlying nature: gravity can be connected to strong interactions in a deep way that we are just beginning to uncover. \par

These models are a good approximation to the study of the excited colored matter created in heavy ion collisions. An interesting next step would be the description of these AdS RN solutions for low temperature, in order to study possible applications to nuclear physics. \par

\textbf{Acknowledgments}: Authors want to thank the Vicedecanatura de Investigaciones of Universidad de los Andes.

\end{document}